Atmospheric Chemistry of Venus-like Exoplanets

by

Laura Schaefer

and

Bruce Fegley, Jr.

**Abstract:** We use thermodynamic calculations to model atmospheric chemistry on terrestrial exoplanets that are hot enough for chemical equilibria between the atmosphere and lithosphere, as on Venus. The results of our calculations place constraints on abundances of spectroscopically observable gases, the surface temperature and pressure, and the mineralogy of the planetary surface. These results will be useful in planning future observations of the atmospheres of terrestrial-sized exoplanets by current and proposed space observatories such as the *Hubble Space Telescope* (HST), *Spitzer*, *James Webb Space Telescope* (JWST), and *Darwin*.

## 1. Introduction

The search for exoplanets in general and Earth-sized exoplanets in particular has been heating up in recent months. Results from the first year of operations of the Kepler mission, which is designed to determine the frequency of Earth-sized exoplanets, are being released (e.g., Borucki et al. 2010) with much more data yet to be analyzed. The COROT space telescope has already made significant discoveries, including one of the smallest and hottest exoplanets so far discovered (CoRoT-7b, Léger et al. 2009). Even ground-based methods have now proven to be capable of detecting super-Earth exoplanets (Charbonneau et al. 2009). Current and future space observatories such as the Hubble Space Telescope (HST), the Spitzer Space Telescope, the James Webb Space Telescope (JWST) or the proposed Darwin mission will also be able to characterize the atmospheres of these exoplanets. As more discoveries of Earth-sized exoplanets are made and characterization of their atmospheres becomes more possible, it is important to model the nature of their atmospheres. What will their main components be? Will they look like the Earth's atmosphere? Or perhaps like the atmospheres of the other terrestrial planets in our own solar system?

Techniques for discovering exoplanets are initially biased towards planets with either short-periods or intermediate periods with high orbital eccentricities (Kane et al. 2009). This is particularly true for transits, in which a planet passes in front of its star and which allow atmospheric observations. As observations for a particular star increase, there is a greater chance of observing longer period planets, but initial discoveries are likely to be of large short period planets (e.g. Borucki et al. 2010). Short-period super-Earth planets like CoRot-7b ($a$ = 0.0172 AU, Léger et al. 2009) should be hot and depleted in volatiles. We have previously modeled such exoplanets, under the assumption that they have been completely stripped of their volatiles (Schaefer & Fegley 2009). Models by others have considered the range of possible compositions we may expect to find for volatile-rich super-Earth exoplanets (e.g., Kaltenegger et al. 2007; Elkins-Tanton & Seager 2008).

In this paper we consider planets that more closely resemble Venus. These are planets that have shorter periods than the star's habitable zone (HZ), and therefore have lost, or never accreted, significant amounts of water. As on Venus, we expect the surface temperature and pressure of these planets to be hot enough to allow surface-atmosphere interactions. Therefore the bulk atmospheric composition will be controlled by the mineralogy of the surface. Models for Venus show that the observed partial pressures of $CO_2$, $H_2O$, HCl, and HF are in chemical equilibrium at a pressure and temperature very close to that observed at the surface of Venus (Fegley 2004; Lewis 1970). In this paper, we apply techniques used to model Venus' atmosphere to models for several hypothetical Venus-like exoplanets.

**2. Venus Surface-Atmosphere Equilibrium Model**

We model atmosphere-lithosphere chemical interactions on exoplanets with surface conditions similar to Venus. We do this by using mineral buffer reactions for minerals that may be plausibly found together in natural rock systems. A mineral buffer reaction is a chemical equilibrium that controls the partial pressure of a gas such as $CO_2$. The partial pressure of a gas is determined solely by the mineral buffer. However, the column density (molecules $cm^{-2}$) also depends on the planet's gravity (and thus size). The intersections of the mineral buffers on a pressure-temperature plot define a set of pressure and temperature conditions for the planet, which allows us to determine within a reasonable range the allowable abundances of $CO_2$, $H_2O$, HCl, and HF that would be present in the atmosphere. This procedure was developed by Lewis (1970) to predict the surface pressure and temperature for Venus. The abundances of $CO_2$, $H_2O$, HCl, and HF measured in the lower Venusian atmosphere allowed Lewis to describe a small suite of possible compatible mineral buffer systems for the surface of Venus. The calculations use plausible mineral buffers – that is reactions involving minerals that are found in the same rock types: felsic rocks with free silica (e.g., like Earth's continental crust), or mafic rocks without free silica (e.g., like Earth's basaltic oceanic crust). We used all buffers considered by Lewis (1970), Fegley and Treiman (1992), and the phyllosilicate buffers considered by Zolotov et al. (1997) in our model (see Appendix 1).

Figure 1a shows results for this method applied to Venus. The point on the graph shows the measured $CO_2$ pressure (taken as the total pressure), and the surface temperature. The lines in Fig. 1a represent the mineral buffers which provide the closest fit to the measured conditions for Venus (T, $P_{CO2}$, $X_{H2O}$, $X_{HCl}$, and $X_{HF}$, where $X_i$ is the mole fraction of gas *i* defined as the partial pressure divided by the total pressure). The model parameters are listed in Table 1, and the mineral buffers used are listed in the appendix. The temperature of the planet is initially defined by the intersections of the $CO_2$ and $H_2O$ buffers, as shown in Fig. 2a. The total pressure of the planet is assumed to be dominated by $CO_2$, so $P_T = P_{CO2}$. The actual surface pressure on Venus is about 95.4 bars due to the presence of ~3.5% $N_2$ in Venus' atmosphere. Neither Lewis' (1970) model nor our calculations can predict the $N_2$ abundance in the atmosphere of a Venus-like planet, which must be inferred by other techniques. However, by analogy with Venus, the amount is presumably small and is neglected to first approximation. The $CO_2$ pressure of Venus (92.1 bars) is most closely matched by the calcite-quartz-wollastonite buffer (reaction C1 in appendix). Our model explicitly assumes that this buffer (C1) controls the $CO_2$ pressure in the lower Venusian atmosphere.

The $H_2O$ mixing ratio is defined by observations and is altitude dependent. The $H_2O$ abundance of the lower atmosphere (0-40 km) is uniform, and is in equilibrium with the surface. Lower atmosphere abundances have been measured by both ground-based and spacecraft observations. A recent review by Fegley (2004) gives an average value of 30±15 ppm for the lower atmosphere. More recent measurements by Venus Express have found abundances of 31±2 ppm (Marq et al. 2008), 44±9 ppm (Bezard et al. 2009), and 22 – 35 ppm (Tsang et al. 2010). Here we adopt the average value of $X_{H2O}$ = 30 ppm. We considered 14 water buffers given by Lewis (1970), Fegley & Treiman (1992), and Zolotov et al. (1997) (see Appendix 1 for a complete list). The intersections of the $CO_2$ and $H_2O$ buffers are shown in Figure 2a for $X_{H2O}$ = 30 ppm, with the point representing Venus surface conditions. The water buffer that intersects the $CO_2$ buffer most closely to the observed T and P conditions (740 K, 92.1 bars) is chosen. This reaction is the eastonite buffer (W1 in appendix 1). The C1 and W1 buffers intersect at 758 K and 122 bars, which is close to the observed surface conditions of Venus.

Figure 2b illustrates how the calculated $H_2O$ abundance in Fig. 1 depends on the surface temperature and pressure. The dark line is the C1 buffer, and the point shows Venus surface conditions. The thinner lines show the change in total pressure with the assumed $H_2O$ mole fraction, from 0.1 ppm to 1% as a function of temperature and pressure. Between 10 and 100 ppm, results are shown for 10 ppm steps. The temperature and pressure for Venus are matched exactly with an $H_2O$ abundance of 24 ppm. This is somewhat lower than the average value of 30 ppm, but well within the ±15 ppm uncertainty (Fegley 2004).

The abundances of HCl and HF depend on both the total pressure and surface temperature, as well as on the $H_2O$ mole fraction. Using the 10 HCl buffers from Lewis (1970) and Fegley & Treiman (1992) (see Appendix 1), we find a range of HCl abundances from 42 ppb – 29 ppm at 740 K, as shown in Figure 3a. For comparison, the observed abundance of HCl in Venus' atmosphere ranges from 0.1-0.2 ppm at 64-94 km (Bertaux et al. 2007) to 0.74 ppm at an altitude of 60-66 km (Iwagami et al. 2008) and 0.40 ppm at 12 – 24 km (Iwagami et al. 2008). Krasnopolsky (2010) found an HCl abundance of 0.40 ppm at 74 km. The HCl abundance is generally considered uniform throughout Venus' atmosphere (Krasnopolsky 2010). Here we adopt a reasonable average value for the HCl abundance of 0.5±0.1 ppm (Fegley, 2004). The best fit for this HCl abundance is given by the albite – halite – andalusite – quartz buffer (Cl1), which gives an HCl abundance of 0.76 ppm.

Figure 3b shows the calculated HF abundance as a function of temperature for the HF buffers listed in Appendix 1. We find a range of HF abundances from 0.2 ppb – 25.7 ppm at 740 K. As with HCl, the HF abundance is generally considered uniform throughout the atmosphere. The Venus Express SPICAV instrument found HF abundances of 1 – 3 ppb at 75 – 85 km (Bertaux et al. 2007). Krasnopolsky (2008, 2010) found an HF abundance of 3.5±0.2 ppb at ~70 km using ground-based observations. We adopt an average value for HF of 4.5 ppb (Fegley 2004). The best fit to the HF abundance is given by the fluor-phlogopite buffer (F1) which gives an HF abundance of 4.6 ppb.

## 3. Models of Lower Atmospheres of Venus-like Exoplanets.

*3.1 Surface-Atmosphere Equilibrium* Figure 1b-1d shows results from our models for 3 hypothetical Venus-like exoplanets. The initial parameters of these models were chosen from intersections of different $CO_2$ and $H_2O$ buffers. These intersections determine the total pressure of $CO_2$ and the temperature. The parameters (T, P, $X_{H2O}$, $X_{HCl}$, $X_{HF}$) for each model and the buffers used are listed in Table 1. The mineral buffer reactions are listed in the appendix.

We explored the necessary conditions to form planets hotter (740 – 1000 K) and colder (450 – 740 K) than Venus for both felsic and mafic mineral suites. We found that in order to create a planet hotter than Venus, the $H_2O$ abundance had to increase significantly. Notice on Figure 2a ($X_{H2O}$ = 30 ppm) that there are few intersections of $CO_2$ and $H_2O$ buffers (other than eastonite, an uncommon mineral) at temperatures to the right (hotter) of the Venus point. As shown in Fig. 2b, increasing the $H_2O$ abundances shifts the lines for the $H_2O$ buffers to the right on the graph, giving more intersections with the $CO_2$ buffers. For Venus-like exoplanets with felsic ($SiO_2$-bearing) crusts like Earth's continental crust, abundances greater than ~100 ppm $H_2O$ were necessary. All hot felsic planets had larger total pressures than Venus. Mafic planets required even larger $H_2O$ abundances (≥ 1000 ppm) for all water buffers other than eastonite, which produced hotter temperatures than Venus for $H_2O$ abundances >30 ppm (see Figure 2b). In general, therefore, higher water vapor abundances should correspond to higher surface temperatures and more mafic surface mineralogies. Mafic planets also produced a much wider range of pressures, some less than and some greater than that of Venus.

The first exoplanet model shown in Fig. 1b (model B in Table 1) is a high-temperature exo-Venus with a basaltic (mafic) mineral suite. The $CO_2$ and $H_2O$ buffers that define the temperature and pressure are reactions C2 (magnesite-enstatite-forsterite) and W2 (phlogopite-forsterite-leucite-kalsilite) in Appendix 1. Using our mafic HCl and HF buffers, we found a range of abundances for HCl (446-544 ppb) and HF (7.4 ppb – 4.34 ppm). In the figure, we show our chosen results for the Cl2 (wollastonite-sodalite-halite-anorthite-albite) and F2 (orthoclase-forsterite-fluorphlogopite-enstatite) buffers, which give 446 ppb HCl and 7.4 ppb HF, respectively. Although we do not show a high temperature felsic planet here, we found that they generally have lower HCl abundances and higher HF abundances than mafic planets. However, the ranges between the two suites overlap significantly, so it is not possible to distinguish between a mafic and felsic mineral suite on this basis alone.

For planets colder than Venus, we found that nearly all mafic exoplanets had lower total surface pressures than Venus, whereas the felsic exoplanets could have pressures both significantly larger and smaller than Venus. As temperature and water vapor abundance increase for the mafic exoplanets, the total pressure increases. Wide ranges of water vapor abundance produced planets colder than Venus for both the felsic and mafic mineral suites. To compare the possible HCl and HF abundances, we chose a cold felsic (model C) and a cold mafic planet (model D) with similar temperatures, pressures, and $H_2O$ abundances. The temperature and pressure of model C are defined by the intersection of the C3 (diopside-quartz-calcite-forsterite) and W3 (tremolite-enstatite-dolomite-quartz) buffers. The temperature and pressure of model D are defined by the C4 (diopside-enstatite-forsterite-dolomite) and W2 (phlogopite-forsterite-leucite-kalsilite)

buffers. Both planets have an $H_2O$ abundance of 100 ppm. The felsic planet (C) has a slightly higher range of HCl (32 ppb–10.98 ppm) and HF (0.1 ppb–12.5 ppm) abundances compared to the mafic exoplanet (4.04-37.2 ppb HCl, 0.12-382 ppb HF). We show representative values in Fig. 1c and 1d. Unfortunately, however, the range of abundances given by possible HCl and HF buffers are not significantly different enough to permit observations to constrain whether an exo-planet's surface is felsic or mafic.

*3.2 Temperature-Pressure Profiles of the Lower Atmosphere* Figure 4 shows the calculated temperature-pressure profiles for our 4 models from 0 to 50 km. The temperature-pressure profile for Venus (model A) is taken from the Venus International Reference Atmosphere (Seiff et al. 1986). The profiles for the exo-Venus models were calculated using the dry adiabatic gradient, assuming the same *g* as Venus, for an atmosphere composed of pure $CO_2$. Notice that the profiles for models C and D, our low temperature felsic and mafic planets, are nearly identical.

The use of a pure $CO_2$ atmosphere is an approximation because the abundance of $N_2$ is unknown. All other gases are assumed to be less abundant than $N_2$ and should not significantly affect the lower atmosphere adiabat. The effect of small amounts of $N_2$ is to increase the overall pressure of the atmosphere, without significantly altering the temperature structure. For an atmosphere of 10% $N_2$ and 90% $CO_2$, we found temperature deviations at 50 km of ~2 K, and an increased pressure of ~22%, amounting to an absolute increase of ~0.2 bars.

A far greater source of error is the value for *g,* which depends upon the size of the exoplanet. Given the current detection limits for exoplanets, any Venus-like exoplanets that are observed are likely to be larger than Venus. A super-Venus is likely to have an atmosphere that is much more compressed than Venus and is therefore vertically shorter, yet significantly denser. An increase of $g$ from 8.87 m $s^{-2}$ to 10 m $s^{-2}$, equal to a planetary radius of 6824 km (~1.13$R_{Venus}$) for a planet with the same density as Venus yields a decrease of ~76% in pressure and ~35% in temperature at an altitude of 50 km. An exo-Venus of ~5$R_{Venus}$ will reach the water vapor condensation point below 10 km. Ehrenreich et al. (2006) have shown that the shorter, denser atmosphere of a super-Venus will make it particularly difficult to detect. They also showed that the atmospheres of smaller planets resembling Venus will be somewhat easier to detect through transit observations.

We do not extend the temperature-pressure profile above 50 km because this is roughly the level of cloud formation on Venus. For larger exoplanets, cloud condensation is likely to occur at significantly lower altitudes. For instance, the nominal "Venus" model of Ehrenreich et al. (2006) has a planetary radius equal to the Earth, giving $g$ of 9.8 m $s^{-2}$ compared to Venus' value of 8.87 m $s^{-2}$. For this planet, with a surface pressure of 100 bars, they calculated that the cloud tops would be at roughly 30 km. The level of cloud condensation will strongly depend upon the atmospheric composition and the UV flux input. The atmosphere above the cloud deck is also significantly altered by photochemistry and stellar heating, making accurate predictions for this area difficult due to the uncertainty in planetary size, orbital distance, stellar input, and abundances of photochemically important gases such as CO and $SO_2$. We discuss the upper atmosphere further in the following sections.

**4. Photochemisty, Clouds, and the Upper Atmosphere**

The composition of Venus' upper atmosphere is determined largely by photochemistry. Several of the gases that we consider in our model ($CO_2$, HCl, and HF), have roughly constant mixing ratios throughout Venus' atmosphere and are not significantly depleted by the photochemistry on Venus. However, the abundance of $H_2O$ is depleted in Venus' upper atmosphere by nearly an order of magnitude ($X_{H2O}$ = 1.2-2.9 ppm at 72 km, Krasnopolsky 2010). Water vapor reacts with photochemically produced $SO_3$ to form $H_2SO_4$, which condenses into a thick, global cloud layer between 45 and 70 km altitude. Table 2 lists the photochemical lifetimes of the major gas species in Venus' atmosphere. We discuss some of the photochemical cycles of the upper atmosphere in the following section, and how these may differ for Venus-like exoplanets.

*4.1 Carbon Dioxide and Monoxide* Carbon dioxide has a nearly constant mixing ratio in the atmosphere of Venus, but is easily converted through photochemistry into CO (see Table 2). However, catalytic cycles reform $CO_2$ from CO + $O_2$ with the reaction

$$CO + OH \dashrightarrow CO_2 + H \quad (1)$$

playing an important role in the Martian and possibly Venusian atmospheres. This reaction or others involving Cl oxides are responsible for regulating the CO abundance and reforming $CO_2$ from CO + $O_2$ on Venus (Yung & DeMore 1999). In the absence of such catalytic cycles, the $CO_2$ in the upper atmosphere of Venus would be completely destroyed in ~14,000 years and all $CO_2$ in the atmosphere within ~5 Myr (Fegley 2004). For an exoplanet depleted in the necessary catalytic gases ($H_2$, Cl, and NO), carbon dioxide may not regenerate from photochemically produced carbon monoxide, and the major gas would shortly become CO.

However, carbon monoxide is also destroyed in Venus' lower atmosphere by reaction with elemental sulfur vapor

$$2CO + S_2 = 2OCS \quad (2)$$

Thus by analogy with Venus we expect that the CO abundances in the atmospheres of Venus-like exoplanets are not simply regulated by mineral buffers, but are instead affected by photochemical production and loss via gas phase catalytic cycles in the stratomesospheres and by thermochemical loss in the near surface tropospheres.

*4.2 Sulfur Dioxide, Water Vapor, and Clouds* Sulfur gases are extremely important in the atmosphere of Venus. Recent measurements from Venus Express give an $SO_2$ abundance in the lower atmosphere of Venus of 130 ± 50 ppm (Marcq et al. 2008). Sulfur dioxide may react with surface minerals, such as $CaCO_3$ to form $CaSO_4$, on relatively fast time scales for high surface temperatures. The observed abundance of $SO_2$ is ~100 times more abundant than predicted from thermodynamics for these mineral-gas reactions, which suggests that $SO_2$ is emitted during volcanic episodes (Fegley et al. 1997). In the absence of an active source, the $SO_2$ in Venus' lower atmosphere would be removed in 1-10 Myr (Fegley & Prinn 1989). In the upper atmosphere, $SO_2$ gas is photochemically converted into $SO_3$, which reacts with water vapor to form $H_2SO_4$ (Krasnopolsky & Pollack 1994) The sulfuric acid condenses to form the very thick clouds on Venus, which extend from ~45 - 70 km. Cloud condensation removes >99 % of the $SO_2$ and >90% of $H_2O$ from the upper atmosphere, leaving an $SO_2$ abundance of 350 ± 50 ppb and an $H_2O$ abundance of 1.2-2.9 ppm at ~72 km (Krasnopolsky 2010).

Bullock and Grinspoon (2001) studied the evolution of climate on Venus, and studied the effect of a variable $SO_2$ abundance on the cloud layer. For larger $SO_2$ abundances, they found that clouds became thicker. Higher $H_2O$ abundances also led to much thicker clouds and higher surface temperatures. For $H_2O$ abundances of ~1000 ppm, they found surface temperatures greater than 900 K. At these temperatures, parts of the surface may begin melting, leading to a partial magma ocean. As $SO_2$ abundances drop, the clouds become thinner. At very low abundances of $SO_2$, the clouds are high, thin $H_2O$ clouds, until even these disappear for an atmosphere with no $SO_2$. In order for an exo-Venus to maintain significant cloud cover, it must therefore have a volcanic source of $SO_2$ gas.

Other sulfur gases, such as $H_2S$, OCS, and $S_2$ - if present in the upper atmospheres - should be photolyzed on fairly short timescales (see Table 2). However, it is more likely that $H_2S$, OCS, and $S_2$ will be regulated by gas phase and gas-solid chemical equilibria in the lower atmospheres of Venus-like exoplanets as is the case on Venus (e.g., see Fegley 2004).

**5. Application to Exoplanets**

We believe that this work is timely because several on-going space missions are searching for Earth-like planets (e.g., Spitzer, HST, COROT, Kepler). However, short-period planets are highly favored by current detection methods (Kane et al. 2009). Transits, which allow transmission spectroscopy of planetary atmospheres, are observed far more frequently for short period planets than long-period planets, and so initial planet detections from these missions should be for short-period planets. Short-period planets are likely to be hot from proximity to their stars and tidal heating (e.g., Jackson et al. 2008a,b). These planets, if similar in size to the terrestrial planets in our own solar system, are more likely to have atmospheres resembling Venus than Earth. These planets will be depleted in water, either from having accreted less of it due to their orbital location, or because they have lost water over time, as Venus is suspected to have done (Fegley, 2004).

*5.1 Lower Atmosphere* Observation of lower atmospheric abundances by transmission spectroscopy of Venus-like exoplanets is likely to be difficult. As previously mentioned, Ehrenreich et al. (2006) have shown that a cloud layer such as the $H_2SO_4$ clouds of Venus is optically thick, which effectively blocks the lower atmosphere and increases the planetary radius observed in transits. Only the atmosphere above the cloud tops would be probed by transmission spectroscopy. Detection of the upper atmosphere of a super-Venus is statistically improbable. Planets smaller than Venus are likely to be more observable, as their atmospheres are less tightly bound.

As with Venus, emission spectroscopy would be necessary to probe the lower atmosphere (i.e., below the cloud deck), which is in equilibrium with the surface. The night-side of Venus has several spectral windows between 1.5 and 2.5 μm that emit thermal radiation from the lower atmosphere and allow Earth-based observations of different levels of the lower atmosphere. These observations have been used to determine the abundances of $H_2O$, HF, HCl, OCS, and CO in Venus' lower atmosphere (e.g., Allen & Crawford 1984; Bézard et al. 1990; de Bergh et al. 1995, Krasnopolsky, 2008, 2010). Thermal emissions from exoplanets have already been observed for several gas giant planets by both Spitzer in the mid-IR and HST in the near-IR using the secondary eclipse technique (Deming et al. 2005; Charbonneau et al. 2005; Grillmair et al. 2007; Richardson et al. 2007; Swain et al. 2008, 2009a,b). These observations have identified a number of molecular species including $H_2O$, $CO_2$, CO, and $CH_4$. The JWST, which will have a greater aperture than Spitzer, will be able to conduct more sensitive observations, which should allow spectroscopic observations of Earth-sized exoplanets, particularly around smaller M-class stars (Clampin et al. 2009). Additionally, several proposed missions (e.g., Terrestrial Planet Finder, Darwin) would use a nulling interferometer, which would block the light of the parent star and image the planet directly (Lawson, 2009; Cockell et al. 2009). Such an instrument would allow better detection of infrared emission from the night-side of a planet, where spectral windows such as those found at Venus may be seen. Therefore, we believe that observations of the lower atmospheres of a super-Venus may be possible in the near future.

*5.2 Upper Atmosphere and Photochemistry around other Stars* Table 3 lists properties of other types of main sequence stars (F2V, K2V, quiescent M) compared to the Sun (G2V) (Kasting et al. 1997; Segura et al. 2003, 2005). The F2V star is larger and hotter than the Sun, whereas the K2V and the M stars are significantly smaller and cooler. The table lists the relative luminosities of these stars and the calculated orbital distance of a planet receiving the same amount of integrated stellar flux as Venus receives from the Sun. Kasting et al. (1997) and Segura et al. (2003, 2005) performed similar calculations for exo-Earths in the habitable zone of other stellar classes, and we use their data for stellar luminosities here. Our calculation neglects the effect of albedo, but still gives a good rough estimate for the location of an exo-Venus around different stellar types. The orbital distances around F and G stars are large enough that it would require several years of observations to detect an exo-Venus via transit methods. Therefore, K and M stars are more likely targets for detecting an exo-Venus.

Table 3 also gives the ratio of the UV part of the spectrum of each stellar class relative to the Sun. For instance, the F2V star, which is 3 times more luminous than the Sun, emits roughly twice the amount of UV radiation as the Sun. Although photoabsorption coefficients, as well as stellar spectra, are wavelength-dependent, this ratio can nonetheless be used to roughly scale the photochemical lifetimes of the major gases as given in Table 2 for planets found around F and K stars. Doing so shows that photochemistry will likely have a more pronounced affect on the atmospheric compositions of Venus-like exoplanets around F stars, and a diminished affect on exoplanets around K stars. The increased UV flux around F stars indicates that $CO_2$ will be more quickly converted to CO in the upper atmosphere. The balance of CO to $CO_2$ will depend upon the abundance of the catalytic gases $H_2$, Cl, and NO. Conversely, Venus-like exoplanets around K stars, which experience lower levels of UV light may have a different atmospheric structure, as photochemistry is necessary to produce the $H_2SO_4$ clouds. These planets may have thinner clouds, or clouds with a different composition, such as pure $H_2O$ clouds. However, the composition of the upper atmosphere is purely speculative, since it is highly dependent on the abundances of gases such as CO and $SO_2$, which are unconstrained by the mineral buffer systems at the surface.

M stars behave very differently from FGK stars. The stellar flux of quiescent M stars is significantly lower than that of the Sun over most wavelengths, and is slightly larger at wavelengths less than ~200 nm (see Fig. 1 of Segura et al. 2005); however, active M stars frequently emit stellar UV and XUV (x-ray and extreme ultraviolet) flares, which increases the UV output by several orders of magnitude. These flares occur frequently, on a time scale of hours to days, leading to a highly variable UV input into the atmospheres of planets in orbit around these stars (Scalo et al. 2007). This will have a significant affect on photochemistry in planetary atmospheres. It seems unlikely that a relatively steady state photochemical cycle, such as observed on Venus for $CO_2$-CO conversion and the formation of $H_2SO_4$ clouds, could be established with such a variable source of radiation.

It has also been shown by Lammer et al. (2007) that coronal mass ejections (CMEs) from M stars may be sufficient to strip a thick atmosphere from an Earth-sized planet within the star's HZ in less than 1 Gyr. Atmospheric stripping is more severe for tidally-locked planets with little to no magnetic moment (such as Venus), although thick $CO_2$ atmospheres survive longer due to lower exobase temperatures (Lammer et al. 2007; Tian 2009). Therefore, in order for an exo-Venus to survive for a significant amount of time around an M star, it must have a magnetic field, unlike Venus itself. Should the atmosphere survive, however, it is likely to look significantly different than that of Venus due to the extreme variability in the UV flux.

## 6. Conclusions

Based on our surface-atmosphere equilibrium model, we can say that planets similar to Venus (i.e., thick $CO_2$ atmospheres with only trace water) are more likely to be colder than Venus rather than hotter. Hotter planets should have significantly more water in their atmospheres, and generally will have higher total pressures. Hot felsic planets will have relatively large pressures and HF abundances, with less water and HCl than a similar mafic planet. Planets colder than Venus are more geochemically plausible. These planets will generally have lower total pressures than Venus and may have water vapor abundances similar or larger than Venus. Cold felsic planets will have higher total pressures, HCl, and HF abundances, but lower $H_2O$ abundances than similar mafic planets.

K stars offer the best opportunity to locate a planet similar to Venus. The orbits around K stars are short enough to allow frequent transit observations, and the decreased UV flux may limit the thickness and opacity of the clouds that can form. The larger luminosity of F stars requires a much larger orbital distance, and the larger UV flux may alter photochemical cycles by depleting the $CO_2$ abundance or generating thicker clouds. M stars have highly variable stellar fluxes, with flares of UV radiation that are likely to disrupt normal photochemical cycles. Coronal mass ejections (CMEs) would strip the planet of atmosphere within 1 Gyr unless it had a significant magnetic field.

A full understanding of the upper atmosphere of a Venus-like exoplanet would require the knowledge of the abundance of trace gases for which there are no good constraints. The upper atmosphere chemistry will also depend heavily on the orbital period and the stellar flux. Predictions for the composition of the lower atmosphere are therefore more robust. However, detection of emissions from the lower atmospheric windows may require the use of a nulling interferometer, such as those proposed for the Darwin and cancelled Terrestrial Planet Finder missions.

**Acknowledgments**

This work is supported by NASA Grant NNG04G157A from the Astrobiology program and NSF Grant AST-0707377. We thank the reviewer for helpful comments.

**References**

Allen, D. A. & Crawford, J. W. 1984. Nature, 307, 222.

Bertaux, J.-L. et al. 2007, Nature, 450, 646.

Bézard, B., de Bergh, C., Crisp, D., & Maillard, J. P. 1990, Nature, 345, 508.

Bézard, B., Tsang, C. C. C., Carlson, R. W., Piccioni, G., Marcq, E., & Drossart, P., 2009, J. Geophys. Res., 114, doi:10.1029/2008JE003251.

Borucki, W. J. et al. 2010, Science, 327, 977.

Bullock, M. A., & Grinspoon, D. H. 2001, Icarus, 150, 19.

Charbonneau, D., et al. 2005. ApJ, 626, 523.

Charbonneau, D. et al. 2009, Nature, 462, 891.

Clampin, M. et al., 2009, Astro2010, Science White Papers, no. 46.

Cockell, C. S. et al. 2009, Astrobiology, 9, 1.

de Bergh, C., Bézard, B., Owen, T., Maillard, J. P., Pollack, J., & Grinspoon, D. 1995, Adv. Space Res., 15, 479.

Deming, D., Seager, S., Richardson, L. J., & Harrington, J. 2005. Nature, 434, 740.

Ehrenreich, D., Tinetti, G., Lecavelier des Etangs, A., Vidal-Madjar, A., & Selsis, F. 2006. A&A, 448, 379.

Elkins-Tanton, L. & Seager, S. 2008. ApJ, 658, 1237.

Fegley, B. Jr. 2004, in Meteorites, Comets, and Planets, ed. A. M. Davis, Vol. 1 Treatise on Geochemistry, ed. K. K. Turekian & H. D. Holland (Oxford: Elsevier-Pergamon), 487.

Fegley, B., Jr., & Prinn, R. G., 1989, Nature, 337, 55.

Fegley, B. Jr., & Treiman, A. H. 1992, in Venus and Mars: Atmospheres, Ionospheres and Solar Wind Interactions, ed. J. G. Luhrmann, M. Tatrallyay, & R. G. Pepin (AGU Geophysical Monograph No. 66), 7.

Fegley, B., Jr., Klingelhofer, G., Lodders, K., & Widemann, T., 1997, in Venus II, ed. S. W. Bougher, D. M. Hunten, & R. J. Phillips (Tucson: Univ. Arizona Press), 591.

Grillmair, C. J., Charbonneau, D., Burrows, A., Armus, L., Stauffer, J., Meadows, V., van Cleve, J., and Levine, D. 2007, ApJ, 658, L115.

Iwagami, N., et al. 2008, Plan. Space Sci., 56, 1424.

Jackson, B., Barnes, R., & Greenberg, R. 2008a, MNRAS, 391, 237.

Jackson, B., Greenberg, R., & Barnes, R. 2008b, ApJ, 681, 1631.

Kaltenegger, L., Traub, W. A. & Jucks, K. W. 2007. ApJ, 658, 598.

Kane, S. R., Mahadevan, S., von Braun, K., Laughlin, G. & Ciardi, D. R. 2009, Publ. Astron. Soc. Pacific 121, 1386.

Kasting, J. F., Whittet, D. C. B., & Sheldon, W. R., 1997, Orig. Life Evol. Biosph., 27, 413.

Krasnopolsky, V. A. 2008, Icarus, 197, 377.

Krasnopolsky, V. A. 2010, Icarus, 208, 539.

Krasnopolsky, V. A., & Pollack, J. B. 1994, Icarus, 109, 58.

Lammer, H., et al. 2007, Astrobio., 7, 185.

Lawson, P. R., 2009, Proc. SPIE, 7440, 744002.

Léger, A. et al. 2009, A&A, 506, 287.

Levine, J. S. 1985, The Photochemistry of Atmospheres: Earth, the Other Planets, and Comets, (London: Academic Press).

Lewis, J. S. 1970, Earth Planet. Sci. Lett., 10, 73.

Marq, E., Bezard, B., Drossart, P., & Piccioni, G. 2008, J. Geophys. Res., 113, doi:10.1029/2008JE003074.

Richardson, L. J., Deming, D., Horning, K., Seager, S., & Harrington, J. 2007, Nature, 445, 892.

Scalo, J., et al. 2007, Astrobio., 7, 85.

Schaefer, L., & Fegley, Jr., B. 2009, Astrophys. J., 703, L113.

Segura, A., Kreelove, K., Kasting, J. F., Sommerlatt, D., Meadows, V., Crisp, D., Cohen, M., & Mlawer, E., 2003, Astrobiol., 3, 689.

Segura, A., Kasting, J. F., Meadows, V., Cohen, M., Scalo, J., Crisp, D., Butler, R. A. H., & Tinetti, G., 2005, Astrobiol., 5, 706.

Seiff, A., et al., 1986, In: Kliore, A. J., Moroz, V. I., Keating, G. M., eds., The Venus International Reference Atmosphere. Pergamon, Oxford, pp. 3-58.

Swain, M. R., Bouwman, J., Akeson, R. L., Lawler, S., & Beichman, C. A. 2008. ApJ, 674, 482.

Swain, M. R., Vasisht, G., Tinetti, G., Bouwman, J., Chen, P., Yung, Y., Deming, D., & Deroo, P. 2009a, ApJ, 690, L114.

Swain, M. R., et al. 2009b, ApJ, 704, 1616.

Tian, F. 2009, ApJ, 703, 905.

Tsang, C. C. et al. 2010, Geophys. Res. Lett. 37, doi:10.1029/2009GL041770.

Yung, Y. L., & DeMoore, W. B. 1999, Photochemistry of Planetary Atmospheres, New York: Oxford University Press.

Yung, Y. L., Liang, M. C., Jiang, X., Shia, R. L., Lee, C., Bezard, B., & Marcq, B. 2009. J. Geophys. Res. 114, doi:10.1029/2008JE003094.

Zolotov, M. Yu., Fegley, Jr., B. & Lodders, K. 1997, Icarus 130, 475.

**Figure Captions**

Figure 1. (a) Best fit of mineral buffer systems to the observed conditions and atmospheric abundances of Venus. The mineral buffer model is then applied to several theoretical exoplanets: (b) hot mafic exo-Venus, (c) cold felsic exo-Venus, (d) cold mafic exo-Venus. The point represents the surface conditions of Venus. Buffer reactions are listed in Table 2.

Figure 2. (a) Intersection of all considered $CO_2$ buffers with all $H_2O$ buffers ($X_{H2O}$ = 30 ppm). (b) Effect of $H_2O$ abundance on total pressure, temperature and $CO_2$ abundance, using the C1 buffer (dark line), and the W1 buffer (thin lines). Lines for the W1 buffer between 10 ppm and 100 ppm are in 10 ppm increments. The point shows the observed surface pressure and temperature of Venus.

Figure 3. (a)HCl abundance in mole fractions as a function of temperature for the considered HCl buffers. The lines are calculated assuming a total pressure of 92.1 bars, and $X_{H2O}$ = 24 ppm. The point shows the measured abundance of HCl in Venus' lower atmosphere (0.5 ppm, reference). The best fit is given by the Cl1 buffer, which gives 0.76 ppm at 740 K. (b) HF abundance in mole fractions as a function of temperature for the considered HF buffers. The lines are calculated assuming a total pressure of 92.1 bars, and $X_{H2O}$ = 24 ppm. The point shows the measured abundance of HF in Venus' lower atmosphere (4.5 ppb, reference). The best fit is given by the F1 buffer, which gives $X_{HF}$ = 4.6 ppb

Figure 4. Temperature – Pressure profiles from 0 – 50 km for the 4 atmospheric models listed in Table 1. The profile for model A (Venus) is the Venus International Reference Atmosphere (VIRA) from Seiff et al. (1986). The profiles for models B, C, and D are calculated using a dry adiabatic gradient for a pure $CO_2$ atmosphere. The points show 10 km altitude increments, with 0 km on the right, and 50 km on the left.

**Table 1.** Model Parameters and Gas Abundances

| Model | T (K) | $P_{CO2}$ (bars) | $X_{H2O}$ (ppm) | $X_{HCl}$ (ppb) | $X_{HF}$ (ppb) | buffers[a] |
|---|---|---|---|---|---|---|
| A (Venus) | 740 | 92.1 | 24 | 760 | 4.6 | C1,W1,Cl1,F1 |
| B (hot mafic) | 790 | 439.4 | 1000 | 446 | 7.4 | C2,W2,Cl2,F2 |
| C (cold felsic) | 647 | 43.3 | 100 | 87 | 1.87 | C3,W3,Cl3,F3 |
| D (cold mafic) | 653 | 41.33 | 100 | 4.04 | 0.13 | C4,W2,Cl4,F2 |

[a]See Appendix 1 for full reaction list.

**Table 2.** Photochemical lifetimes at zero optical depth (top of atmosphere) for major atmospheric gases

| Species | $J_1$ ($s^{-1}$) 1 AU[a] | 0.72 AU | $t_{chem}$ (s)[b] |
|---|---|---|---|
| $CO_2$ | $2.02\times10^{-6}$ | $3.90\times10^{-6}$ | $2.56\times10^{5}$ |
| CO | $6.459\times10^{-7}$ | $1.25\times10^{-6}$ | $8.00\times10^{5}$ |
| $SO_2$ | $2.491\times10^{-4}$ | $4.81\times10^{-4}$ | $2.08\times10^{3}$ |
| OCS[c] | $1.97\times10^{-5}$ | $3.81\times10^{-5}$ | $2.62\times10^{4}$ |
| HCl | $7.2\times10^{-6}$ | $1.39\times10^{-5}$ | $7.19\times10^{4}$ |
| HF | $1.8\times10^{-6}$ | $3.47\times10^{-6}$ | $2.88\times10^{5}$ |
| $H_2O$ | $11.8038\times10^{-6}$ | $2.28\times10^{-5}$ | $4.39\times10^{4}$ |

[a]Levine (1985). [b]at 0.72 AU. [c]Yung et al. (2009)

**Table 3.** Stellar Properties and calculated Orbital Distance of Venus around Other Stars

| Stellar Class | $T_{eff}$ (K) | $L/L_{\odot}$ | UV flux 200-400 nm | $a$ (AU)[a] |
|---|---|---|---|---|
| F2V[b] | 6930 | 3.0 | 2.10 | 1.2 |
| G2V (Sun)[b] | 5780 | 1.0 | 1.00 | 0.7 |
| K2V[b] | 4780 | 0.27 | 0.40 | 0.4 |
| M (AD Leo)[c] | 3400 | $2.3\times10^{-2}$ | $3.4\times10^{-3}$ | 0.13 |

[a] orbital distance of a planet receiving the same total amount of stellar flux as Venus around other star types. [b]Kasting et al. (1997). [c]*AD Leo* - Segura et al. (2005).

| Appendix 1 Buffers used in Calculations | |
|---|---|
| *$CO_2$ buffers* | |
| C1 | $CaCO_3 + SiO_2 = CaSiO_3 + CO_2$ |
| C2 | $MgCO_3 + MgSiO_3 = Mg_2SiO_4 + CO_2$ |
| C3 | $2CaMg(CO_3)_2 + SiO_2 = 2CaCO_3 + Mg_2SiO_4 + 2CO_2$ |
| C4 | $CaMg(CO_3)_2 + 4MgSiO_3 = 2Mg_2SiO_4 + CaMgSi_2O_6 + 2CO_2$ |
| C5 | $CaCO_3 = CaO + CO_2$ |
| C6 | $MgCO_3 = MgO + CO_2$ |
| C7 | $FeCO_3 = FeO + CO_2$ |
| C8 | $MgCO_3 + SiO_2 = MgSiO_3 + CO_2$ |
| C9 | $FeCO_3 + SiO_2 = FeSiO_3 + CO_2$ |
| C10 | $CaCO_3 + MgSiO_3 = CaMgSiO_4 + CO_2$ |
| C11 | $2MgCO_3 + SiO_2 = Mg_2SiO_4 + 2CO_2$ |
| C12 | $CaMg(CO_3)_2 + 2SiO_2 = CaMgSi_2O_6 + 2CO_2$ |
| C13 | $CaMg(CO_3)_2 + SiO_2 = CaCO_3 + Mg_2SiO_4 + CO_2$ |
| C14 | $CaMg(CO_3)_2 = CaCO_3 + MgO + CO_2$ |
| C15 | $CaCO_3 + Mg_2SiO_4 = CaMgSiO_4 + MgO + CO_2$ |
| C16 | $CaCO_3 + CaMgSi_2O_6 = Ca_2MgSi_2O_7 + CO_2$ |
| C17 | $2CaCO_3 + CaMgSi_2O_6 + Mg_2SiO_4 = 3CaMgSiO_4 + CO_2$ |
| C18 | $CaCO_3 + MgSiO_3 + SiO_2 = CaMgSi_2O_6 + CO_2$ |
| C19 | $CaCO_3 + Ca_3Si_2O_7 = 2Ca_2SiO_4 + CO_2$ |
| C20 | $2CaCO_3 + CaMgSi_2O_6 = Ca_3MgSi_2O_8 + 2CO_2$ |
| *Water vapor buffers* | |
| W1 | $KMg_2Al_3Si_2O_{10}(OH)_2 = MgAl_2O_4 + MgSiO_3 + KAlSiO_4 + H_2O$ |
| W2 | $2KMg_3AlSi_3O_{10}(OH)_2 = 3Mg_2SiO_4 + KAlSi_2O_6 + KAlSiO_4 + 2H_2O$ |
| W3 | $Ca_2Mg_5Si_8O_{22}(OH)_2 = 3MgSiO_3 + 2CaMgSi_2O_6 + SiO_2 + H_2O$ |
| W4 | $KAl_3Si_3O_{10}(OH)_2 + SiO_2 = KAlSi_3O_8 + Al_2SiO_5 + H_2O$ |
| W5 | $KAl_3Si_3O_{10}(OH)_2 = KAlSi_3O_8 + Al_2O_3 + H_2O$ |
| W6 | $Mg_3Si_4O_{10}(OH)_2 = 3MgSiO_3 + SiO_2 + 2H_2O$ |
| W7 | $5Mg_3Si_2O_5(OH)_4 = Mg_3Si_4O_{10}(OH)_2 + 6Mg_2SiO_4 + 9H_2O$ |
| W8 | $Mg(OH)_2 = MgO + H_2O$ |
| W9 | $Mg(OH)_2 + SiO_2 = MgSiO_3 + H_2O$ |
| W10 | $Mg_3Si_2O_5(OH)_4 + Mg(OH)_2 = 2Mg_2SiO_4 + 3H_2O$ |
| W11 | $Al_2O_2(OH)_2 = Al_2O_3 + H_2O$ |
| W12 | $KMg_3AlSi_3O_{10}(OH)_2 + 3MgSiO_3 = 3Mg_2SiO_4 + KAlSi_3O_8 + 2H_2O$ |
| W13 | $KMg_3AlSi_3O_{10}(OH)_2 + 3SiO_2 = 3MgSiO_3 + KAlSi_3O_8 + 2H_2O$ |
| W14 | $2NaCa_2Mg_4Al_3Si_6O_{22}(OH)_2 = CaAl_2Si_2O_8 + 2NaAlSiO_4 + 3CaMgSi_2O_6 + 2Mg_2SiO_4 + MgAl_2O_4 + 2H_2O$ |
| *Chlorine Buffers* | |
| Cl1 | $2HCl + 2NaAlSi_2O_6 = 2NaCl + Al_2SiO_5 + 3SiO_2 + H_2O$ |
| Cl2 | $12HCl + 6CaSiO_3 + 5Na_4[AlSiO_4]_3Cl = 17NaCl + 6CaAl_2Si_2O_8 + 3NaAlSi_3O_8 + 6H_2O$ |
| Cl3 | $2HCl + 8NaAlSi_3O_8 = 2Na_4[AlSi_3O_8]_3Cl + Al_2SiO_5 + 5SiO_2 + H_2O$ |
| Cl4 | $2HCl + 9NaAlSiO_4 = Al_2O_3 + NaAlSi_3O_8 + 2Na_4[AlSiO_4]_3Cl + H_2O$ |
| Cl5 | $2HCl + 2NaAlSiO_4 + CaSiO_3 = 2NaCl + CaAl_2Si_2O_8 + SiO_2 + H_2O$ |
| Cl6 | $2HCl + 2NaAlSi_3O_8 = 2NaCl + Al_2SiO_5 + 5SiO_2 + H_2O$ |
| Cl7 | $2HCl + 6NaAlSiO_4 + 2NaAlSi_3O_8 = 2Na_4Al_3Si_3O_{12}Cl + Al_2SiO_5 + 5SiO_2 + H_2O$ |
| Cl8 | $2HCl + 8NaAlSiO_4 = 2Na_4Al_3Si_3O_{12}Cl + Al_2SiO_5 + SiO_2 + H_2O$ |
| Cl9 | $6HCl + 2Na_4Al_3Si_3O_{12}Cl + 3CaSiO_3 = 8NaCl + 3CaAl_2Si_2O_8 + 3SiO_2 + 3H_2O$ |
| | *Fluorine buffers* |
| F1 | $2 HF + KAlSi_2O_6 + 2Mg_2SiO_4 = KMg_3AlSi_3O_{10}F_2 + MgSiO_3 + H_2O$ |
| F2 | $2HF + KAlSi_3O_8 + 3Mg_2SiO_4 = KMg_3AlSi_3O_{10}F_2 + 3MgSiO_3 + H_2O$ |
| F3 | $2HF + NaAlSiO_4 + 2CaMgSi_2O_6 + 3MgSiO_3 = NaCa_2Mg_5Si_7AlO_{22}F_2 + SiO_2 + H_2O$ |
| F4 | $CaF_2 + SiO_2 + H_2O = CaSiO_3 + 2HF$ |
| F5 | $2CaF_2 + SiO_2 + MgSiO_3 + 2H_2O = Ca_2MgSi_2O_7 + 4HF$ |
| F6 | $CaF_2 + MgSiO_3 + 2H_2O = CaMgSiO_4 + 2HF$ |
| F7 | $MgF_2 + SiO_2 + H_2O = MgSiO_3 + 2HF$ |
| F8 | $MgF_2 + MgSiO_3 + H_2O = Mg_2SiO_4 + 2HF$ |
| F9 | $Na_3AlF_6 + 10SiO_2 + Ca_2Al_2SiO_7 + 3H_2O = 3NaAlSi_3O_8 + 2CaSiO_3 + 6HF$ |
| F10 | $2HF + KAlSi_2O_6 + 3MgSiO_3 = KMg_3AlSi_3O_{10}F_2 + 2SiO_2 + H_2O$ |
| F11 | $2HF + KAlSi_3O_8 + 3MgSiO_3 = KMg_3AlSi_3O_{10}F_2 + 3SiO_2 + H_2O$ |
| F12 | $2HF + NaAlSiO_4 + 2CaMgSi_2O_6 + Mg_2SiO_4 + MgSiO_3 = NaCa_2Mg_5Si_7AlO_{22}F_2 + H_2O$ |
| F13 | $2HF + SiO_2 + CaMgSi_2O_6 + 3MgSiO_3 = Ca_2Mg_5Si_8O_{22}F_2 + H_2O$ |
| F14 | $2HF + 2CaMgSi_2O_6 + 5MgSiO_3 = Ca_2Mg_5Si_8O_{22}F_2 + Mg_2SiO_4 + H_2O$ |
| F15 | $4HF + CaAl_2Si_2O_8 + 2NaAlSiO_4 + 3CaMgSi_2O_6 + 2Mg_2SiO_4 + MgAl_2O_4 = 2NaCa_2Mg_4Al_3Si_6O_{22}F_2 + 2H_2O$ |

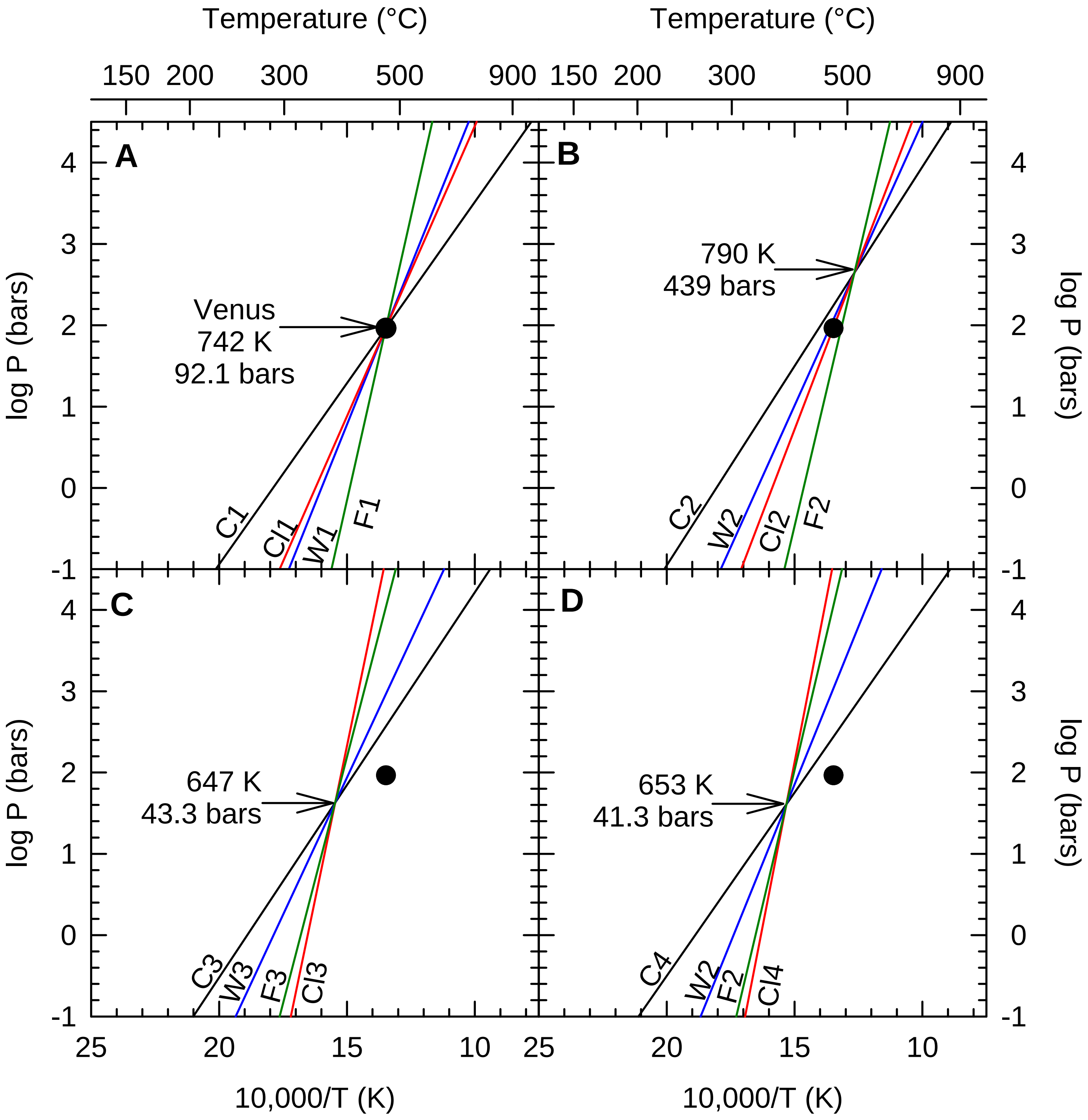

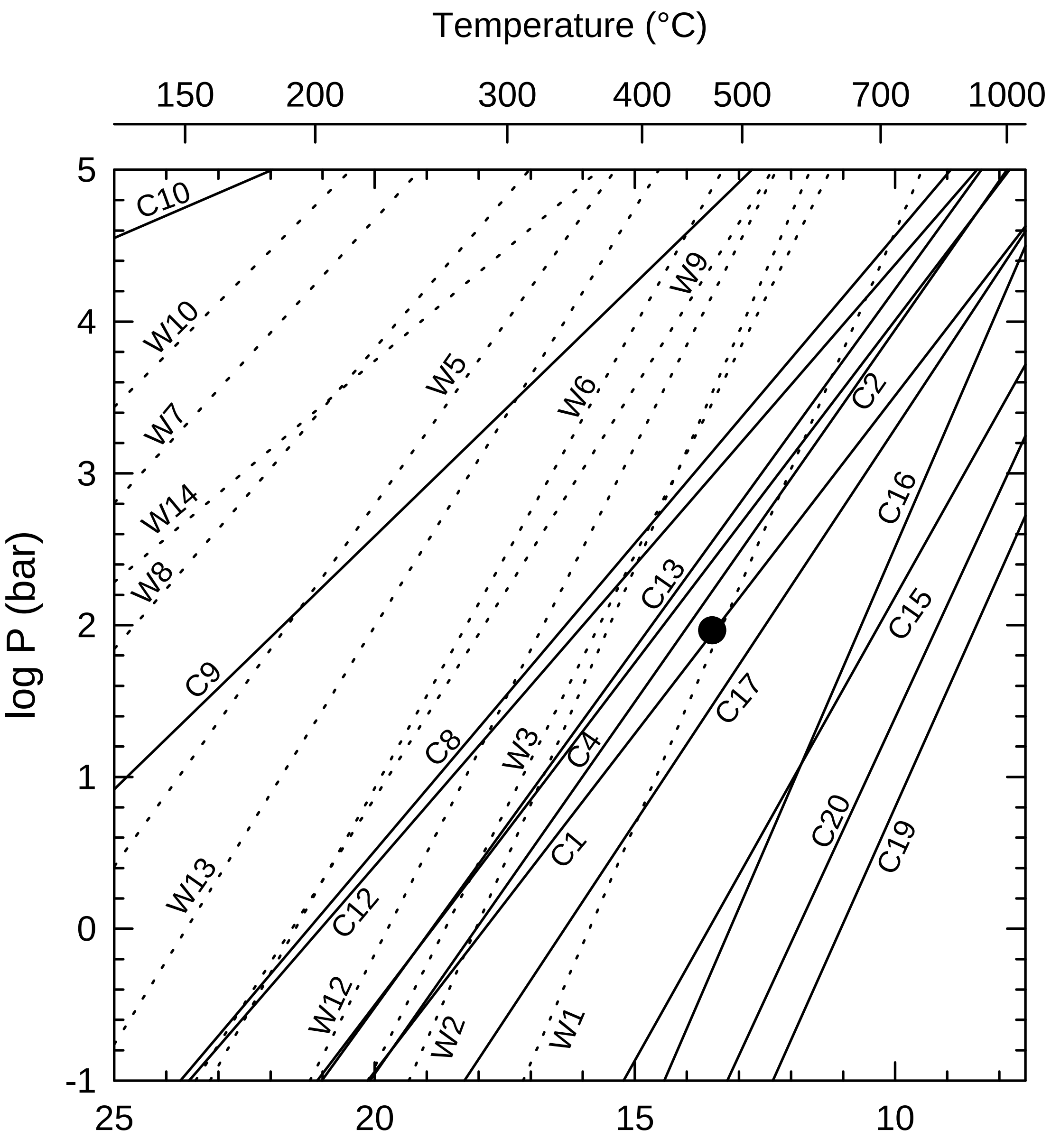

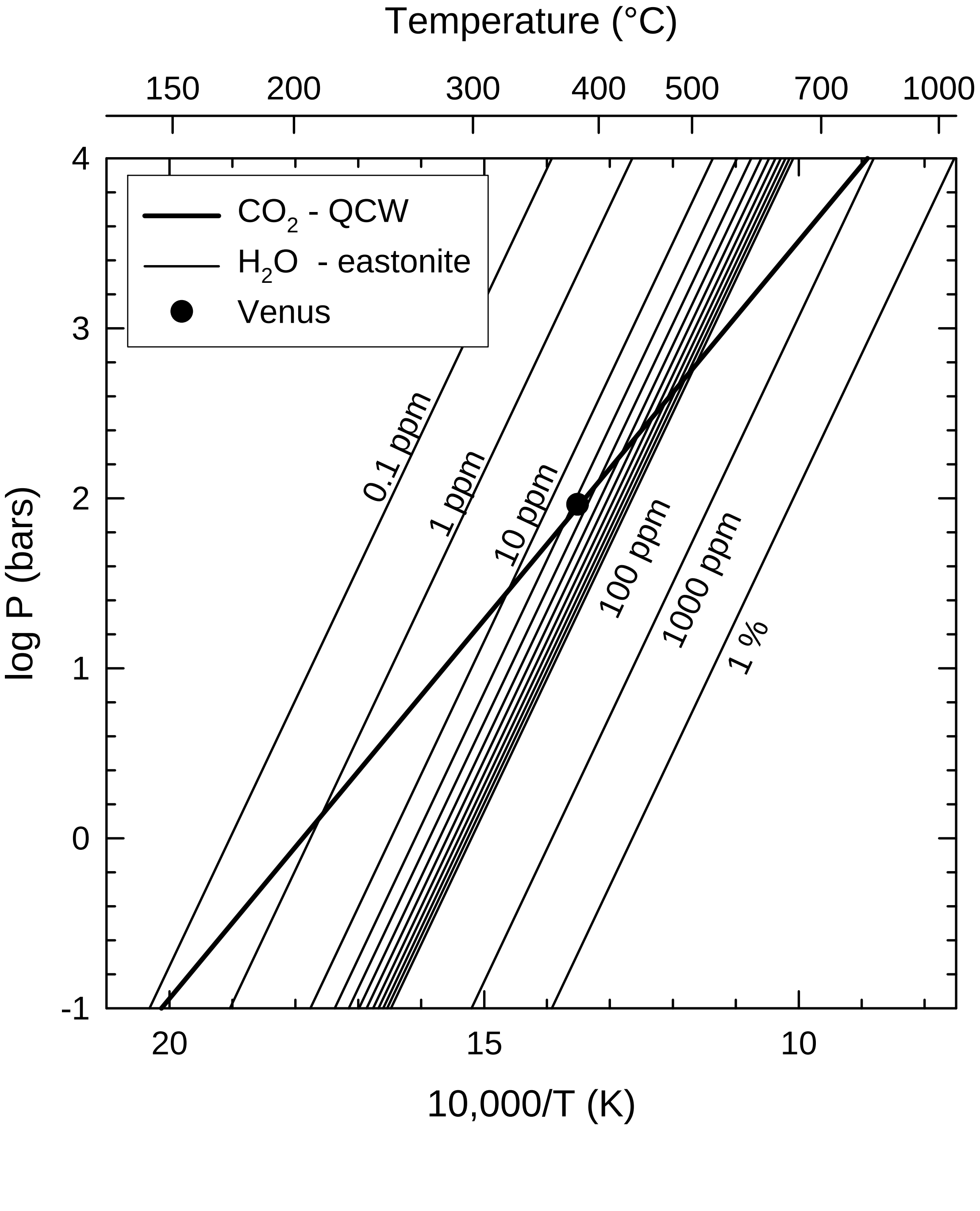

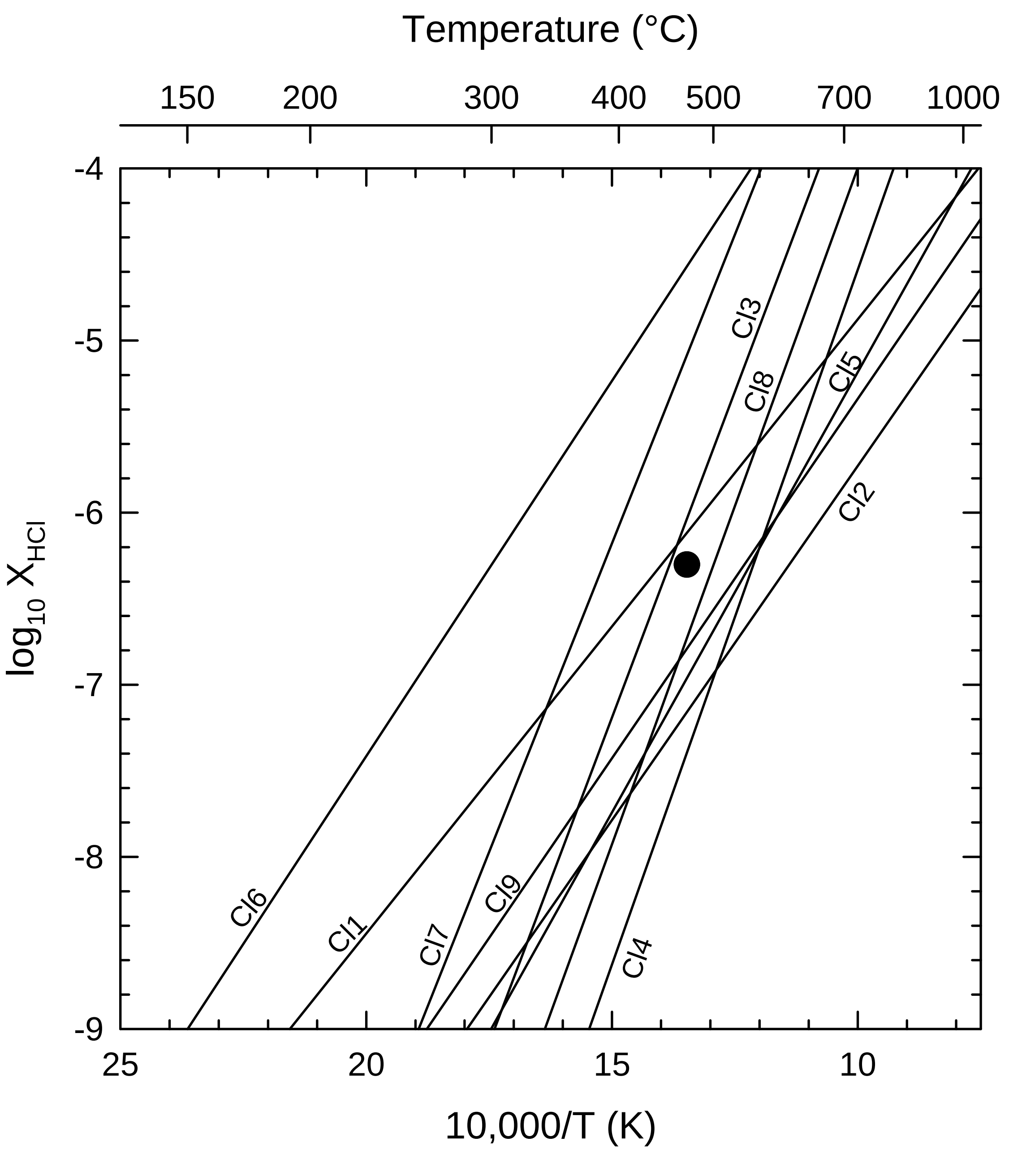

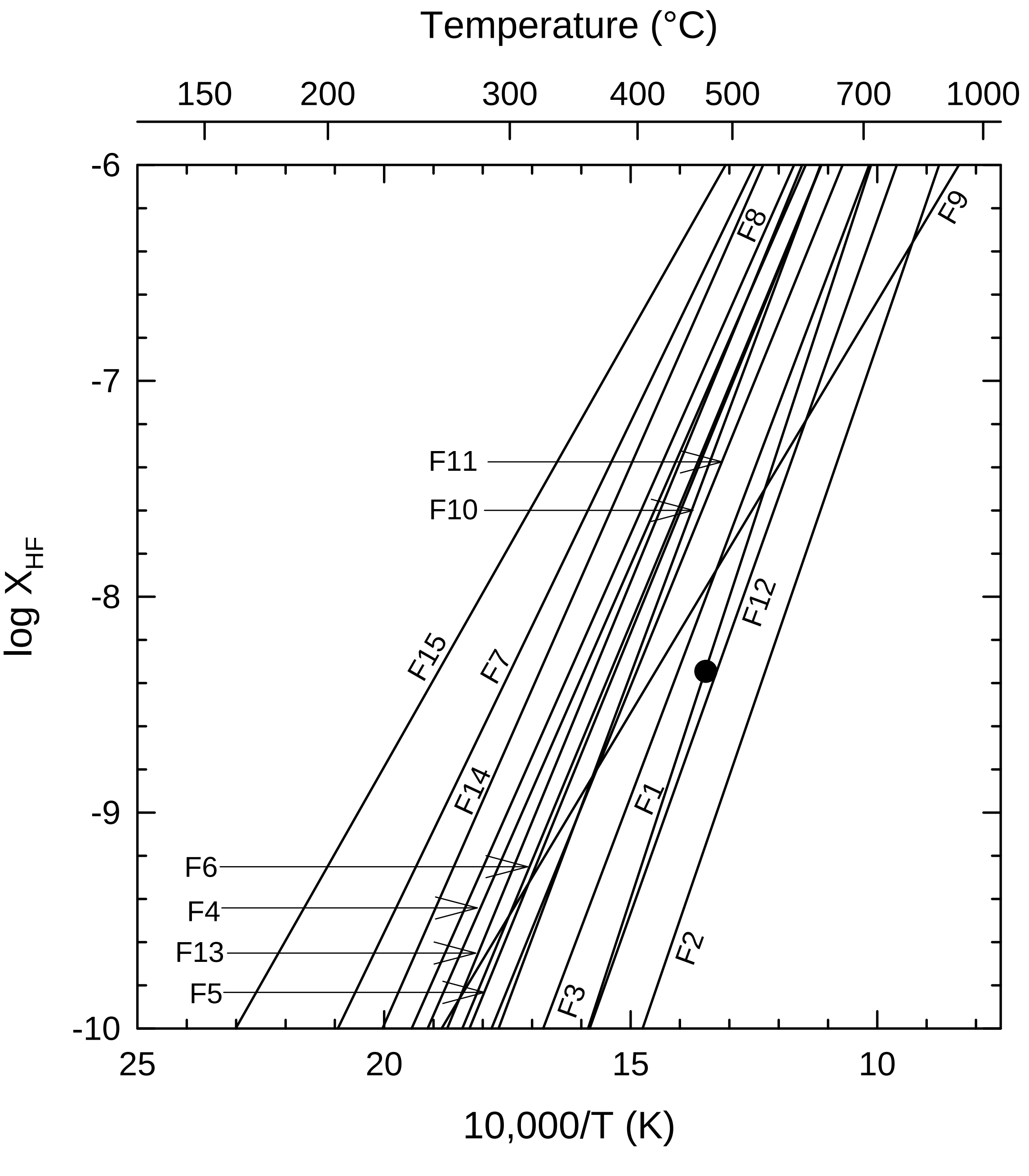

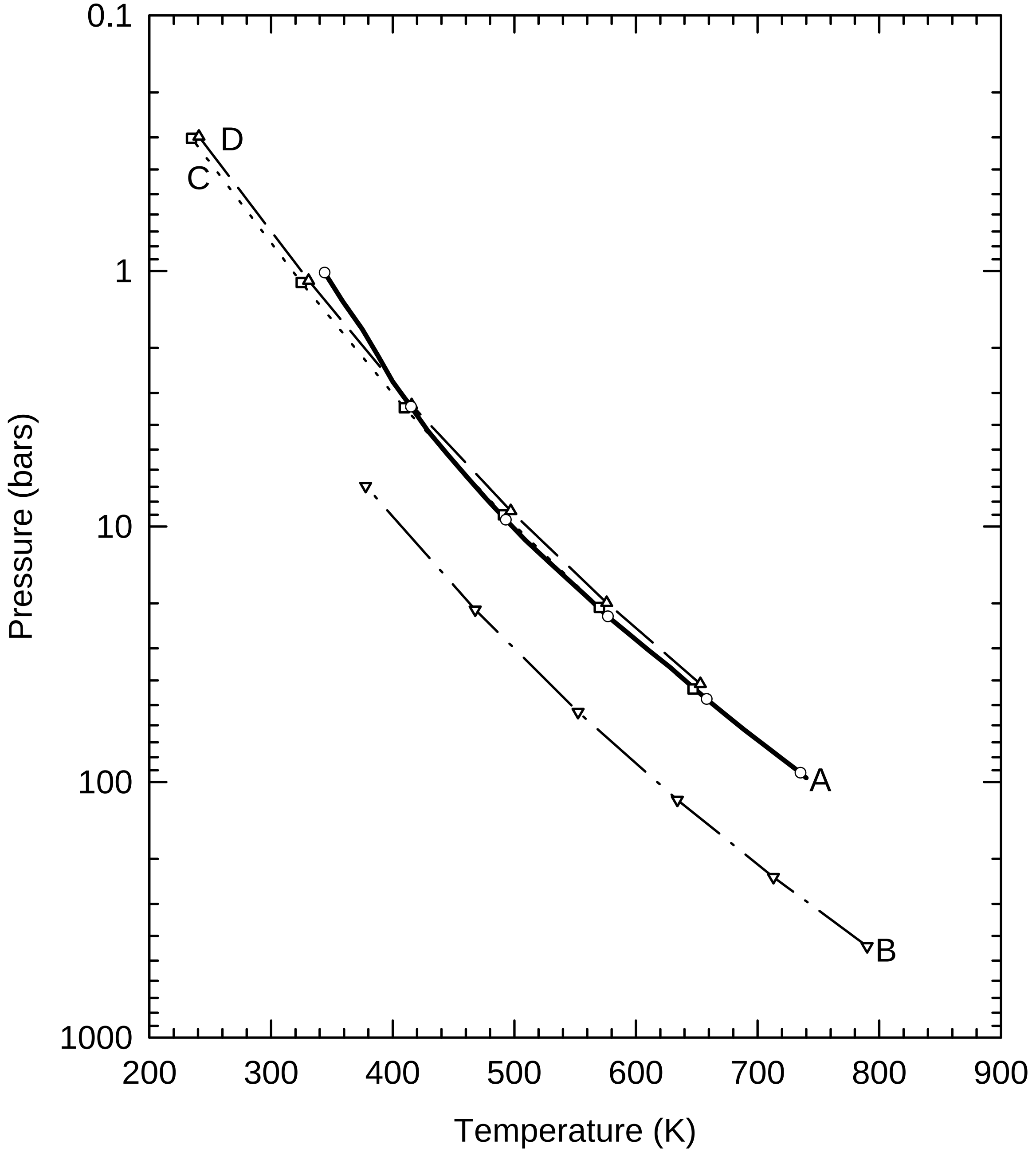